\documentstyle[12pt]{article}
\begin{document}
\begin{center}

{\large
{Generalized Fock Spaces and New Forms of Quantum Statistics}
}

\vskip 1.2cm

{\bf A. K. Mishra $^{*\dagger}$ and G. Rajasekaran $^{\dagger}$ }

\vskip 0.5cm
{\it {$^*$ Max-Planck Institute for Physics of Complex Systems,  
Nothnitzer Str. 38, D-01187 Dresden, Germany, and  \\
$^{\dagger}$ Institute of Mathematical Sciences,
CIT Campus, Madras - 600 113, India\\
e-mail: mishra@imsc.ernet.in; graj@imsc.ernet.in}}

\end{center}

\vskip 1.2cm

\baselineskip=18pt
\centerline{\bf Abstract}

\vskip 0.5cm

\noindent
The recent discoveries of new forms of quantum
statistics require a close look at the under-lying Fock space structure.
This exercise becomes all the more important in order to provide a
general classification scheme for various forms of statistics, and
establish interconnections among them whenever it is possible. We
formulate a theory of generalized Fock spaces, which has a three tired 
structure consisting of Fock space, statistics and algebra. This
general formalism unifies various forms of statistics and algebras,
which were earlier considered to describe different systems. Besides,
the formalism allows us to construct many new kinds of quantum
statistics and the associated algebras of creation and destruction
operators. Some of these are: orthostatistics, null statistics or
statistics of frozen order, quantum group based statistics and its
many ${\it avatars}$, and `doubly-infinite' statistics. The emergence
of new forms of quantum statistics for particles interacting with
singular potential is also highlighted.

\newpage

\baselineskip=20pt

\section*{1. Introduction }

 In recent years, many new kinds of quantum statistics have been
postulated. In spite of large literature which now exists, a 
unified picture for various statistics and their associated algebra
has not emerged. The aim of the present work is to provide such a 
general formalism. This is achieved by introducing the 
concept  of generalized Fock spaces.
Starting with this basic notion,
it is possible to show that more than one statistics can be postulated 
in a given Fock space, and many different  algebraic realizations
can be constructed for any particular statistics.

We introduce new forms of quantum
statistics, {\it viz.} null, orthofermi and Hubbard statistics, and 
doubly-infinite statistics in the next section.
The null statistics corresponds to a situation wherein no permutation 
is allowed and particles are frozen in their initial order. Orthofermi
and Hubbard statistics satisfy an exclusion principle which is
more exclusive than the Pauli's exclusion principle: an orbital
state shall not contain more than one particle irrespective of
their spin directions. Such a situation arises when the Coulomb
repulsion $U$  between two electrons occupying a same orbital state becomes
infinity. The $U$ infinity model has been extensively used in the context
of strongly correlated electron systems [1]. A deformation of the
orthofermi algebra subsequently leads to doubly-infinite statistics.

 The theory of generalized Fock spaces is formulated in  Sec.3. The key
element is the notion of independence of the permutation ordered
states. The largest linear vector space constructed in this way
is the super Fock space. The subsequent specification of a subset
of states in this space as null states leads to many reduced Fock
spaces. All these spaces are collectively called as generalized Fock
spaces. We construct creation ($c^{\dagger}$), annihilation ($c$)
and number ($N$) operators
in the generalized Fock spaces. The creation and annihilation
operators, even for a particular Fock space, are not unique.
Consequently, many statistics and algebras can exist in a given
Fock space. On the other hand, a universal representation for
the number operator valid for all forms of statistics and
algebra exists.

 In quantum mechanical calculations, algebraic relations involving
only $c$ and $c^{\dagger}$ are required. One does not explicitely
need a $c~c$ relation. However, given a $c~c^{\dagger}$ expression,
corresponding $c~c$ relation can be obtained in an elegant manner.
This is demonstrated in Sec.4, and based on this approach, fractional
statistics in one dimension is constructed. Sec.5 is devoted to 
summary and conclusions.

\section*{2. New Forms of Quantum Statistics }
Newer forms of quantum statistics have been constructed by deforming
the canonical commutation relations. For example, the deformation 
$$
[c_j, c^{\dagger}_k]_{^-_+}  =
\delta_{j k} \quad  \rightarrow   \quad 
[c_j, c^{\dagger}_k]_q  =  c_j c_k^{\dagger}  + q c_k^{\dagger} c_j = 
\delta_{j k}  \quad ; \quad -1 < q < 1 
\eqno(1)
$$
gives rise to infinite statistics [2,3]. Here no $c~c$ relation exists, and
all permuted states are linearly independent. 

As a counterpoint to infinite
statistics, null statistics can also be constructed [4]. As mentioned 
earlier, no permutation is allowed in the null statistics. The defining algebra
for this statistics is :
$$
c_k c^\dagger_j \ = \ 0 \quad {\rm for} \quad k \ne j ~~;~~ 
c_j c^\dagger_j \ = \ 1 - \sum_{k < j} c^\dagger_k c_k ~~;~~
c_i c_j \ = \ 0 \quad {\rm for} \quad i < j \eqno(2)
$$

Singular interparticle interactions also lead to new kinds of statistics.
When in addition to Pauli's exclusion principle, an infinite repulsion
exists between two particles occupying
the same orbital state (k or m) but having different spin directions
($\alpha$ and $\beta$), we have
$$
c_{k \alpha} c_{k \beta} \ = \ 0   \eqno(3)
$$
For usual fermions, the above relation is valid only when $\alpha~ = ~
\beta$. Consistent with the above exclusion principle, two
different statistics, $\it viz.$ orthofermi statistics
 $$
c_{k\alpha} c^\dagger_{m\beta} + \delta_{\alpha\beta} \sum_\gamma
c^\dagger_{m\gamma} c_{k\gamma} \ = \ \delta_{km} 
\delta_{\alpha\beta} \quad ; \quad 
c_{k\alpha} c_{m\beta} + c_{m\alpha} c_{k\beta} \ = \ 0\,.
\eqno(4)
$$ 
and Hubbard statistics
$$
\left. \begin{array}{c}
c_{k\alpha} c^\dagger_{m\beta} + (1-\delta_{km})
c^\dagger_{m\beta} c_{k\alpha} \ = \ \delta_{km} \delta_{\alpha\beta}
\left( 1 - \sum_\gamma c^\dagger_{k\gamma} c_{k\gamma}\right) 
\\
\\
c_{k\alpha} c_{m\beta} + (1-\delta_{km}) c_{m\beta} c_{k\alpha}
\ = \ 0
\end{array} \right\} 
\eqno(5)
$$ 
can be postulated [5]. The  orthofermi statistics is formulated in
a representation invariant manner. The Hubbard statistics is not invariant
under unitary transformation, and it depends on the representation. 

The algebra for the orthobose statistics is obtained by replacing 
the positive signs by  negative signs in Eq.(4).

Usually states of a system are  characterized by a set of individual 
indices describing position, spin, internal degrees of freedom etc.. These are 
then mapped to a set of single indices. The symmetry properties are
then postulated with respect to these composite indices. As a result,
symmetries with respect to the exchange of individual indices get correlated.
For the ortho and Hubbard statistics, spatial and spin indices can not be 
mapped to a single composite index. But for Hubbard statistics, exchange
between $k\alpha$ and $m\beta$ is  still permissible. In orthostatistics,
exchanges between $k$, $m$  and between $\alpha$, $\beta$ are uncorrelated. The former
exchange leads to fermi or bose  statistics, where as the later satisfies 
infinite statistics. 

A deformation of  $c$ $c^{\dagger}$ algebra for the 
orthostatistics, that is, 
$$
c_{k\alpha} c^\dagger_{m\beta} + q \delta_{\alpha\beta} \sum_\gamma
c^\dagger_{m\gamma} c_{k\gamma} \ = \ \delta_{km} 
\delta_{\alpha\beta}\quad = \quad -1 < q < 1 
\eqno(6)
$$ 
gives rise to doubly-infinite statistics; one with respect to 
the  pair of indices $(k, m)$
and  the other for the pair $(\alpha, \beta)$.

\section*{3. Generalized Fock Spaces}
Given a set of quantum numbers $g, h, i... $ with the respective occupancy
being $n_g, n_h, n_i...$, all possible multiparticle state vectors are
$$ 
\vert n_g, n_h \ldots n_m ; \mu \rangle \quad , \quad \mu = {
1,2} \ldots s      
\eqno(7)
$$ 
where $s$ is the total number of distinct permutations and $\mu$ labels each
of these permuted states.
we assume the existence of a unique vacuum state
$$
\vert 0 \rangle \equiv \vert 0, 0, 0 \ldots 0 \rangle \eqno(8)   
$$
All these states are linearly independent, but need not be orthogonal or
normalized
$$
\langle n^{'}_g, n^{'}_h \ldots n^{'}_m ; \mu \vert n_g,n_h \ldots n_m ;
\nu \rangle \, = \, \delta_{n^{'}_g n_g} \, \delta_{n^{'}_h n_h}
\ldots \delta_{n^{'}_m n_m} \, M_{\mu \nu} \eqno(9)     
$$
with $M$ being a $s \times s$ hermitian matrix. We choose it to be positive 
definite.

From the set of linearly independent state vectors, an orthonormal set
of vectors 
$\{\parallel n_g \ldots n_m ; \mu \gg \} $
can be obtained
$$
\parallel n_g \ldots n_m ; \mu \gg \, =\, \sum_\nu X_{\nu \mu} \vert n_g
\ldots n_m ; \nu \rangle    
\eqno(10)
$$ 

Alternatively, starting with the orthonormal vectors
$\{\parallel n_g \ldots n_m ; \mu \gg \}$, the vectors
$\{\vert n_g, n_h \ldots n_m ; \mu \rangle \} $  can be constructed by taking 
the inverse of relation (10).
$X$ is a nonsingular matrix. Although $X$ is not unique and depends on the 
particular orthogonalization procedure, we have
$$
M^{-1} \,=\, XX^{\dagger}     
\eqno(11)
$$ 

Choosing a nonsingular matrix $X$ and determining 
the inner product matrix $M$
as above will ensure
the positivity of the matrix $M$.

The set of state vectors considered here constitute super Fock space.
Infinite statistics resides in this Fock space.  Using the projection
operator
$$
P(n_g \ldots n_k
\ldots n_m)       
= \sum_{\lambda, \nu} \, \vert n_g \ldots n_m ; \nu \rangle
(M^{-1})_{\nu \lambda} \, \langle n_g \ldots n_m ; \lambda \vert
\eqno(12)
$$
the number operator can be written as
$$
N_k \, =\, \sum_{n_g \ldots n_k \ldots n_m} \quad n_k P(n_g \ldots n_k
\ldots n_m)       
\eqno(13)
$$ 
which satisfies the following properties
$$
N_k \vert n_g \ldots n_k \ldots n_m ; \mu \rangle \, =\, n_k \vert n_g
\ldots n_k \ldots n_m ; \mu \rangle  \quad  ; \quad 
[N_k, N_j]_- \, = \, 0     
\eqno(14)
$$

The creation operator is defined as
$$
c^{\dagger}_j \,=\, \sum_{n_g \ldots n_j \ldots n_m} \, \sum_{\mu
\nu} \, A_{\mu \nu} \, \vert 1_j n_g \ldots n_j \ldots n_m ; \mu
\rangle \, \langle n_g \ldots n_j \dots n_m ; \nu \vert    
\eqno(15)
$$ 
and $c_j$ as the hermitian conjugate of $c_j^{\dagger}$.
$A_{\mu \nu}$ are a set of arbitrary (complex) numbers. Even at this stage,
it is possible to verify that
$$
[c^{\dagger}_j, N_k ]_- \, =\, - c^{\dagger}_j \, \delta_{jk}    
\eqno(16)
$$ 

The ordered state vectors can be constructed using a string of $c^{\dagger}$
acting on the vacuum state. Consequently we also have
$$
\sum_{\nu} \, A_{\mu \nu} \, M_{\nu \lambda} \,=\, \delta_{\mu
\lambda} \quad ; \quad     
A \,=\, M^{-1}     
\eqno(17)
$$ 

We have provided here a unique representation of the number operator.
But many different representations of creation and annihilation operators
are possible through different choices of matrices $A$, $X$ and $M$. 
The number operator
$N$ can be expressed in terms of $c^{\dagger}$ and $c$ by solving
Eqs.(12,13) and (15). Since $c^{\dagger}$ and $c$ are not uniquely defined
many different expressions for $N$ in terms of $c^{\dagger}$, $c$ can be
obtained.

Next we consider reduced Fock spaces.  
All known forms of statistics other than the infinite 
statistics reside in reduced Fock spaces, which are obtained by postulating
relations like
$$
\sum_{\mu} B^p_\mu |n_g,n_h \ldots ; \mu > = 0 \ ; \ p = 1,2, \ldots r  
\eqno(18)
$$ 
where
r $ < $ s and  $B^p_{\mu}$ are constants. The vector space dimension in the  
sector $\{n_g, n_h \ldots
\}$ is now reduced to  $d = s - r$. 
The formalism developed for the super Fock space is also valid for reduced
Fock spaces. But $\mu$ and $\nu$ now ranges from 1...d, and $X$, $M$ and $A$
are $d \times d$ matrices. 
 
Arbitrariness in the matrix $A$ appearing in the creation operator expression
(15) can be exploited to generate many relations involving $c^{\dagger}$
$c^{\dagger}$. Therefore,  many different forms of statistics specified by 
different $c^{\dagger}$
$c^{\dagger}$ relations can be constructed in a 
given Fock space. All these statistics are interconnected. No connection 
exists between the statistics and the associated algebras residing in different 
Fock spaces. It may also be mentioned here that multiplicity of 
statistics are not possible in the  super Fock space and in the 
Fock space of frozen order. Only infinite and null statistics respectively
reside in these two Fock spaces. But even here, many algebras involving
$c$ and $c^{\dagger}$ are possible. For example, depending on the different 
choice of the the inner product matrix $M$ in the super Fock space, new
$c$ $c^{\dagger}$ relations in addition to the one given in Eq.(1) are
possible. Some of these are ($p > $ 1 or $p < $ -1):
$$
c_ic^\dagger_j -  q \delta_{ij} \sum_k
c^\dagger_k c_k \ = \ \delta_{ij}  \quad ; \quad -1 < q < \infty 
\eqno(19)
$$
$$
c_i c^\dagger_j - c^\dagger_j c_i \ = \ \delta_{ij} p^{2\sum_{k<i}N_k}
p^{N_i}    
\eqno(20)
$$
$$
c_i c^\dagger_j - p^{-1} c^\dagger_j c_i \ = 0 \quad {\rm for} \quad i \ne
j \quad ; \quad  
c_i c^\dagger_i - c^\dagger_i c_i \ = \ p^{N_i}  
\eqno(21)
$$ 

The
$q$ in Eq.(1) can be made a complex number, provided the indices are ordered.
$$
c_i c^\dagger_j - q c^\dagger_j c_i = 0 \quad {\rm for} \quad i < j
\eqno(22)
$$
For completeness this relation has to be supplemented with (p real)
$$
c_i c^\dagger_i - p c^\dagger_i c_i = 1   
\eqno(23)
$$ 
$p$ = $\vert  q \vert $ corresponds to infinite statistics when 
$\vert q \vert$ $ < $ 1.
$p$ = -1 and $\vert q \vert$ $ < $ 1
leads to infinite statistics with an exclusion principle.

Similarly, many algebraic relations can be obtained in the bosonic
Fock space ($d$ = 1). 
Taking q and p to be  complex numbers and $\phi$ as any arbitrary
function of number operator,
a general $c_i$ $c_i^{\dagger}$ relation in this space is written as
$$ 
c_j c_j^\dagger - p c^\dagger_j c_j \ = \ |q|^{2 \sum_{i<j} N_i} f(N_j)
\quad ; \quad
f(N_j) \ = \ \left \vert \frac{\phi (N_j)}{\phi(N_j+1)} \right \vert^2
-p \left \vert \frac{\phi(N_j-1)}{\phi(N_j)} \right \vert^2 
\eqno(24)
$$
Most interestingly, the corresponding 
$c_i$ $c_j^{\dagger}$ relation for i $ < $ j is still given by
Eq.(22). Thus replacing Eq.(23) by Eq.(24) takes us from super
Fock space and infinite statistics to bosonic Fock space and deformed
bose statistics satisfying the symmetry relation
$$
c_i^\dagger  c_j^\dagger - q c_j^\dagger  c_i^\dagger \, = 0 \quad 
{\rm for}  \quad i > j    
\eqno(25)
$$

Various limiting cases, like $\phi$ being a constant, or p = 0, or
$f$ = 1 and $p$ real are possible for Eq.(24). A particular interesting
case is $p$ = $\vert q \vert^2$. The complete algebra (Eqs.(22,24,25))
now becomes covariant under $SU_q(n)$ or quantum group transformation [6-8]. 
This shows that the algebra covariant under quantum group is a particular
case of the more general algebra that can be derived from the formalism
of generalized Fock spaces.

The canonical Bose statistics as well as the q-bose statistics given 
by relation (25) reside in the bosonic Fock space and are interconnected [8]. 
The underlying configuration space for the q-statistics is 
non-commutative. It has been shown that a complete Fock space realization
of the differential calculus on a non-commutative space leads to
a new concept of simultaneous transmutation between quanta satisfying
different quantum statistics [9].

Restricting state occupancy to zero and one in the bosonic Fock
space leads to the fermionic Fock space. Note that no restriction is
placed on the symmetry properties of the state vectors. Consequently,
it is possible to construct anticommuting bosons in the bosonic
Fock space and commuting fermions in the fermionic Fock space.
In addition, many distinct algebras can be obtained in the fermionic
Fock space too.  

A detailed list of statistics and algebras in 
various Fock spaces corresponding to single, and two-indexed systems 
(e.g., orthofermi, and particles obeying `doubly-infinite' statistics)
are provided in reference [8].

\section*{4. $c c$ relations from $c c^{\dagger}$ algebra}
A general form of $c$ $c^{\dagger}$ algebra which allows us to 
to calculate  vacuum matrix element of any polynomial in $c$ and 
$c^{\dagger}$ is given as 
$$
c_i c^\dagger_j \ = \ A_{ij} + \sum_{k,m} B_{ijkm} c^\dagger_k c_m
\eqno(26)
$$ 
where  $A_{ij}$ and  $B_{ijkm}$ are constants or functions of number
operators.

Symmetry of particles under exchange is obtained by making the operator
$Q_{ij}  \equiv 
c_i c_j - q^{'} c_j c_i  $ 
a null operator ($Q_{ij} = 0$). This can be achieved if it can be shown that
$$
Q_{ij} c^{\dagger}_k \,=\, \sum_{i'j'k'} \, F_{ijk; i'j'k'} \,
c^{\dagger}_{k'} \, Q_{i'j'}   
\eqno(27)
$$
for all i,j and k where
F$_{ijk;i'j'k'}$ may be a c-number or operator. The successive applications
of the above equation over any string of creation operators, and then 
allowing both side of the resulting expression to act on the vacuum state
$\vert 0 \rangle $ finally leads to the operator identity
$
Q_{ij} \, = \, c_i c_j - q^{'} c_j c_i   \,=\, 0
$, 
which is the $c$ $c$ relation sought after.
 
Employing the above methodology with the $c$ $c^{\dagger}$ relations 
given in Eqs.(22,23), we can show that fractional statistics with (without)
exclusion principle occurs in one dimension  when $q = e^{i\theta}$
and  $ p = -1~ (\ne -1)$ [10]. For other values of $q$, no $c$ $c$ relation
exists. This provides an analytical method to prove the absence  of a $c$ $c$
relation for infinite statistics.

\section*{5. Summary and Conclusions}
By decoupling the notion of  the underlying Fock space from $c$ and
$c^{\dagger}$, we are able to define different forms of statistics in
a representation independent manner. Subsequently, one can construct
creation, annihilation operators and their algebra in any desired representation.

The  general formalism not only unifies and classifies various forms 
of quantum statistics, but also enables us to construct many new kinds
of statistics and algebras for single and two-indexed systems in a systematic
manner. Some of these are: (i) null statistics, 
(ii) orthostatistics, (iii) doubly-infinite
statistics, (iv) complex q or fractional statistics in one dimension. Many  
$c$ $c^{\dagger}$ algebras representing these statistics are also constructed.
Besides, the notion of generalized Fock
space leads to the concept of statistical transmutation in a quantum
plane.

\end{document}